\documentclass[a4paper,12pt]{article}

\usepackage[english]{babel}
\usepackage[utf8x]{inputenc}
\usepackage[T1]{fontenc}
\usepackage{authblk}
\usepackage[dvipdfmx]{graphicx}
\usepackage{natbib}
\usepackage{here}
\usepackage{setspace}
\usepackage{multirow}
\usepackage{lmodern}
\usepackage{amsmath, amssymb}
\usepackage{bm}
\usepackage{pdflscape}
\usepackage{authblk}
\usepackage{color}
\usepackage[table]{xcolor}

\usepackage{soul}

\usepackage{booktabs}
\usepackage{afterpage}
\usepackage{textcomp}
\usepackage{float}
\usepackage{enumerate}
\usepackage{rotating}

\definecolor{babyblue}{rgb}{0.54, 0.81, 0.94}
\definecolor{babypink}{rgb}{0.96, 0.76, 0.76}

\usepackage[margin=1in]{geometry}

\usepackage{amsmath}
\usepackage{graphicx}
\usepackage[colorlinks=true, allcolors=blue]{hyperref}

\usepackage{caption}
\usepackage{subcaption}

\begin{document}

    \begin{center}
        \vspace*{1cm}
        \large
        \textbf{Integrated community occupancy models: A framework to assess occurrence and biodiversity dynamics using multiple data sources}\\
         \normalsize
           \vspace{5mm}
         Jeffrey W. Doser\textsuperscript{1, 2}, Wendy Leuenberger\textsuperscript{2, 3}, T. Scott Sillett\textsuperscript{4}, Michael T. Hallworth\textsuperscript{5}, Elise F. Zipkin\textsuperscript{2, 3}
         \vspace{5mm}
    \end{center}
    \small
         \textsuperscript{1}Department of Forestry, Michigan State University, East Lansing, MI, USA \\
         \textsuperscript{2}Ecology, Evolution, and Behavior Program, Michigan State University, East Lansing, MI, USA \\
         \textsuperscript{3}Department of Integrative Biology, Michigan State University, East Lansing, MI, USA \\
         \textsuperscript{4}Migratory Bird Center, Smithsonian Conservation Biology Institute, National Zoological Park, Washington, DC, USA \\
         \textsuperscript{5}Vermont Center for Ecostudies, Norwich, VT, USA \\
         \noindent \textbf{Corresponding Author}: Jeffrey W. Doser, email: doserjef@msu.edu; ORCID ID: 0000-0002-8950-9895 \\
          \noindent \textbf{Running Title}: Integrated community occupancy models
         
\section*{Abstract}

\begin{enumerate}
    \item The occurrence and distributions of wildlife populations and communities are shifting as a result of global changes. To evaluate whether these shifts are negatively impacting biodiversity processes, it is critical to monitor the status, trends, and effects of environmental variables on entire communities. However, modeling the dynamics of multiple species simultaneously can require large amounts of diverse data, and few modeling approaches exist to simultaneously provide species and community level inferences.
    \item We present an ``integrated community occupancy model'' (ICOM) that unites principles of data integration and hierarchical community modeling in a single framework to provide inferences on species-specific and community occurrence dynamics using multiple data sources. The ICOM combines replicated and nonreplicated detection-nondetection data sources using a hierarchical framework that explicitly accounts for different detection and sampling processes across data sources. We use simulations to compare the ICOM to previously developed hierarchical community occupancy models and single species integrated distribution models. We then apply our model to assess the occurrence and biodiversity dynamics of foliage-gleaning birds in the White Mountain National Forest in the northeastern USA from 2010-2018 using three independent data sources.
    \item Simulations reveal that integrating multiple data sources in the ICOM increased precision and accuracy of species and community level inferences compared to single data source models, although benefits of integration were dependent on data source quality (e.g., amount of replication). Compared to single species models, the ICOM yielded more precise species-level estimates. Within our case study, the ICOM had the highest out-of-sample predictive performance compared to single species models and models that used only a subset of the three data sources. 
    \item The ICOM provides more precise estimates of occurrence dynamics compared to multi-species models using single data sources or integrated single-species models. We further found that the ICOM had improved predictive performance across a broad region of interest with an empirical case study of forest birds. The ICOM offers an attractive approach to estimate species and biodiversity dynamics, which is additionally valuable to inform management objectives of both individual species and their broader communities.
\end{enumerate}

\noindent \textbf{Keywords}: avian, Bayesian, data fusion, data integration, hierarchical modeling, imperfect detection, joint likelihood

\newpage 

\section*{Introduction}

Populations and communities of a wide range of organisms, including birds \citep{rosenberg2019decline}, bats \citep{rodhouse2019evidence}, and insects \citep{wagner2021insect}, have experienced severe declines in their distributions and abundances across large geographical regions as a result of habitat loss, climate change, and other anthropogenic stressors. As a result, there is growing interest in developing enhanced monitoring techniques to estimate species' trends and distributions. While species-level assessments can be valuable, it is critical to quantify the status and dynamics of entire communities to  understand global change effects on biodiversity. However, multi-species approaches require large numbers of observations for each species or make assumptions about equal detectability of species during sampling, precluding robust assessments of rare species and overall community dynamics \citep{sor2017effects}. 

Despite these limitations, many sampling approaches (e.g., autonomous recording units for birds and bats, camera traps for mammals) naturally provide data on the occurrence patterns of multiple species in a community. However, occurrence data for each species are confounded by imperfect detection during sampling. An observer may record a species as absent if it is indeed absent, or the observer may fail to detect the species during sampling, and thus we refer to such data as detection-nondetection data. Replicated sampling can provide additional information on species' detection rates, allowing for the separation of the occurrence process from the detection process to enable ecologically relevant inference. Most commonly, this information is obtained by collecting multi-species detection-nondetection data on several occasions over short time periods when closure---no permanent extinction or colonization---can be reasonably assumed \citep{mackenzie2005designing}. By accounting for imperfect detection of species during sampling, these data (hereafter ``replicated data'') allow for estimation of species distributions and occurrence patterns within an occupancy modeling framework \citep{mackenzie2002estimating}. This additional information required to separate detection from occurrence is often more costly and time consuming to collect, which motivates the desire to use nonreplicated detection-nondetection data to model species distribution patterns. Accordingly, there is widespread interest in determining statistically robust ways to incorporate nonreplicated data sources into species and community level analyses to inform effective biodiversity conservation.  

Two recent statistical modeling developments, data integration and hierarchical community occupancy models, have led to improved inference on individual species and community dynamics, respectively, to help inform biodiversity assessments. Data integration is a model-based approach to combining multiple data types \citep{zipkin2018synthesizing, miller2019recent}. This approach yields many key benefits compared to single data source models, such as higher accuracy and precision of parameter estimates \citep{dorazio2014accounting}, inference across broader spatio-temporal extents \citep{zipkin2021addressing}, and the opportunity to accommodate sampling biases and imperfect detection for the various data sources \citep{miller2019recent}. Data integration is particularly useful for combining large-scale nonreplicated data with replicated data, as the replicated data allow for the explicit modeling of imperfect detection while still using the information contained in the nonreplicated data. The combination of these data types can thus greatly expand the spatial scope of analysis. 

Hierarchical community occupancy models provide combined inference on the occurrence of multiple species using a single replicated detection-nondetection data source \citep{dorazio2005estimating, gelfand2005modelling}. Species-specific parameters are viewed as random effects arising from a common, community-level distribution with a mean and variance parameter representing the average effect across species in the community and the variation of species-specific effects within the community, respectively \citep{dorazio2005estimating}. In addition to providing more precise estimates of species-specific effects \citep{zipkin2009impacts}, these models can estimate community-level parameters, as well as biodiversity metrics \citep{guillera2019inferring}, all with associated uncertainty \citep{zipkin2010multi}. Community occupancy models have led to improved insights into understanding species occurrence patterns for insects \citep{mata2017conserving}, birds \citep{zipkin2009impacts}, and mammals \citep{gallo2017mammal}.

While substantial development of both data integration and hierarchical community models has occurred over the last decade (reviewed in \citealt{zipkin2018synthesizing, guillera2019inferring, miller2019recent}), there has been comparatively less research focused on modeling multiple species using more than one data source. Multi-species integrated population models have been used to assess competition \citep{peron2012integrated}, synchrony \citep{lahoz2017bringing}, and predator-prey dynamics \citep{barraquand2019integrating, clark2021large}, but these approaches are not designed for assessment of larger communities. \cite{clark2017generalized} introduced a broad framework (GJAM) for using multiple data sets to estimate the distribution and abundance of multiple species, but their approach does not model the observation process hierarchically, making it difficult to account for differences in sampling protocols and species detections among data sources. Thus, a new modeling approach that combines the benefits of data integration and hierarchical community modeling in a single framework has the potential to yield detailed inferences on individual species occurrence patterns while simultaneously providing inference on community dynamics.

Here we develop an ``integrated community occupancy model'' to simultaneously estimate occurrence patterns of multiple species within a community as well as community metrics by incorporating multiple available data sources into a single analysis. Our modeling framework combines one or more ``replicated'' detection-nondetection data sources with one or more ``nonreplicated'' data sources. The integrated model consists of a single, ecological process model and multiple observation models that explicitly account for different detection processes in each data source. The detection models are linked to the process model using a joint-likelihood framework \citep{miller2019recent}. Our ecological process model is a dynamic community occupancy model in which we explicitly model the latent occurrence process as a function of space and/or time varying covariates and a temporal autologistic parameter \citep{royle2008hierarchical}. We validate our model using simulations and assess its ability to estimate species-specific and community level dynamics using different amounts of data sources. We then compare inferences from our integrated community occupancy model to those generated by single species integrated models to evaluate the marginal benefits of the approach. We apply our model to an empirical case study of a community of twelve foliage-gleaning insectivorous birds in the White Mountain National Forest in the northeastern USA to assess patterns and trends in species occurrence and community metrics (i.e., richness, composition) from 2010-2018. Our integrated community occupancy model provides a rigorous approach to elucidate both species specific and community level dynamics that can provide crucial insight on the specific mechanisms driving biodiversity shifts and population declines.

\section*{Materials and Methods}

\subsection*{Integrated community occupancy model}

We develop an ``integrated community occupancy model'' (ICOM) that leverages one or more replicated detection-nondetection data sources with one or more nonreplicated detection-nondetection data sources to provide inferences on species-specific and community occurrence dynamics. We present the model using one replicated and one nonreplicated data source, but our framework can be extended to incorporate additional data sources if available in an analogous manner. The ICOM consists of a single ecological process model for individual species, which is shared across the various data sets. Here, we demonstrate the model with a dynamic ecological process that describes species occurrences as a function of spatio-temporally varying covariates and an autologistic parameter that accounts for temporal correlation \citep{royle2008hierarchical}. This process model is then linked to individual likelihoods for each data set via a hierarchical framework that assumes independence among the detection processes and which is conditional on the true latent occurrence state (i.e., whether a species is truly present or absent at sampling locations; \citealt{miller2019recent}). To link the individual species models, we assume species-level occurrence and detection parameters are random effects coming from common community level distributions \citep{dorazio2005estimating, gelfand2005modelling}, enabling information sharing across the community to increase precision of species effects and estimation of biodiversity attributes. 

\subsubsection*{Ecological Process Model}

Our goal is to model the occurrence dynamics of multiple species at sites $j = 1, \dots J$ within a specified region of interest $A$. Let $z_{i, j, t}$ denote the true presence (1) or absence (0) of species $i$ at site $j$ during year $t$, where $i = 1, \dots, I$ and $t = 1, \dots, T$. We assume $z_{i, j, t}$ arises from a Bernoulli process following 

\begin{equation}\label{initOcc}
  z_{i, j, t} \sim \text{Bernoulli}(\psi_{i, j, t}),
\end{equation}

where $\psi_{i, j, t}$ is the probability species $i$ occurs at site $j$ in year $t$. For the first year, we model $\psi_{i, j, t}$ according to 

\begin{equation}\label{psi1}
  \text{logit}(\psi_{i, j, 1}) = \beta0_{i, 1} + \bm{\beta}_{i} \cdot \bm{x}_{j, 1}, 
\end{equation}

where $\beta0_{i, 1}$ is the species-specific occurrence probability (on the logit scale) in the first year (at average covariate values) and $\bm{\beta}_i$ is a vector of species-specific regression coefficients that describe the effect of standardized covariates (i.e., mean 0 and standard deviation 1) $\bm{x}_{j, 1}$ on the occurrence probability of species $i$. In subsequent years, the occurrence probability for species $i$ in year $t$ at site $j$ depends on whether or not the species was present at the site $j$ in the previous year $t - 1$ in addition to covariates (which can vary spatially and/or temporally). We accommodate the temporal dependence by incorporating a species-specific autologistic parameter $\phi_i$ into the occurrence model, such that for $t > 1$

\begin{equation}\label{psit}
  \text{logit}(\psi_{i, j, t}) = \beta0_{i, t} + \bm{\beta}_{i} \cdot \bm{x}_{j, t} + \phi_{i} \cdot z_{i, j, t -1},  
\end{equation}

where $\beta0_{i, t} + \phi_{i}$ is the species-specific intercept in year $t$ when species $i$ occurred at site $j$ in the previous year $t - 1$ and $\beta0_{i, t}$ is the intercept in year $t$ when species $i$ did not occur at site $j$ in the previous year $t - 1$. We use the autologistic parameterization of the dynamic community occupancy model as it allows us to assess covariate effects directly on species occurrence probabilities (i.e., the covariate effects remain the same regardless of the value of $z_{i, j, t - 1}$) and we can derive species-specific trends post-hoc from the occurrence probabilities ($\psi_{i, j, t}$). However, the ecological process model can be readily modified to incorporate relevant biological processes of interest (provided sufficient data are available; \citealt{royle2008hierarchical}). For example, the ecological process model can be modified to explicitly include a trend effect (covariate on year), assess covariate effects on colonization and persistence \citep{dorazio2010models}, or the autologisitc component can be removed if that is not relevant to the target community. 

\subsubsection*{Observation Model: replicated detection-nondetection data}

For the replicated data type, we assume $K > 1$ ``sampling replicates'' within each year $t$ are available at a subset of sites $r = 1, \dots, R$. The sampling replicates can be observations from multiple independent surveys, multiple independent observers, spatial subsamples, or sampling intervals from a removal design, which enable separate estimation of occurrence and detection probability \citep{mackenzie2002estimating}. We assume the $R$ sites are a subset of the total $J$ sites (i.e., $R \leq J$), which may cover the entire region of interest $A$ or, more commonly, only a portion of it. Let $y_{i, r, k, t}$ denote the detection (1) or nondetection (0) of species $i$ during replicate $k$ at site $r$ during year $t$.  We model $y_{i, r, k, t}$ as

\begin{align}
    y_{i, r, k, t} &\sim \text{Bernoulli}(p_{i, r, k, t} \cdot z_{i, j[r], t}),
\end{align}

where $p_{i, r, k, t}$ is the probability of detecting species $i$ during visit $k$ at site $r$ in year $t$ and $z_{i, j[r], t}$ is the true occurrence status of species $i$ in year $t$ at site $j$ corresponding to the $r$th replicated data site.

Species detection probabilities can vary by site and/or sampling covariates following

\begin{equation}\label{det-p}
   \text{logit}(p_{i, r, k, t}) = \alpha0_{i, t} + \bm{\alpha}_{i}  \cdot \bm{w}_{r, k, t} 
\end{equation}

where $\alpha0_{i, t}$ is the species-specific detection probability (on the logit scale) in year $t$ at average covariate values and $\bm{\alpha}_{i}$ is a vector of parameters that describe the effect of standardized covariates $\bm{w}_{r, k, t}$ on the detection probability of species $i$. 

\subsubsection*{Observation Model: nonreplicated detection-nondetection data}

Let $v_{i, m, t}$ be the detection (1) or non-detection (0) of species $i$ at site $m$ in year $t$ for the nonreplicated data source, where $m = 1, \dots, M$. We assume the $M$ sites are a subset of all $J$ sites of interest (i.e., $M \leq J$) within the area of interest $A$. The replicated data may be available at different sites than the nonreplicated data in the same region $A$, the same sites, or a subset of the same sites in $A$. 

We model the detection-nondetection data $v_{i, m, t}$ according to 

\begin{align}
  v_{i, m, t} &\sim \text{Bernoulli}(\pi_{i, m, t} \cdot z_{i, j[m], t}), \label{pa1}
\end{align}

where $\pi_{i, m, t}$ is the probability of detecting species $i$ in site $m$ in year $t$ for the nonreplicated data set and $z_{i, j[m], t}$ is the true occurrence status of species $i$ in year $t$ at the site $j$ corresponding to the $m$th nonreplicated data site.  Detection probability $\pi_{i, m, t}$ can vary by species, site, and time following

\begin{equation}\label{det-pi}
  \text{logit}(\pi_{i, m, t}) = \gamma0_{i, t} + \bm{\gamma}_i \cdot \bm{s}_{m, t},
\end{equation}

where $\gamma0_{i, t}$ is a species and year specific intercept and $\bm{\gamma}_i$ is a vector of parameters that describe the effect of standardized covariates $\bm{s}_{m, t}$ on the detection probability of species $i$. 

Nonreplicated data alone are unable to separate occurrence probabilities from detection probabilities as the model structure is generally unidentifiable \citep{dorazio2014accounting}. While detection and occurrence parameters can be weakly identifiable under certain circumstances (e.g., detection varies with covariates, large number of sites and years; \citealt{lele2012dealing, keryRoyle2020}), estimates often do not converge or have unreasonably large credible intervals, such that inferences are not ecologically useful. 

\subsubsection*{Linking species models across the community}

Following the structure of the hierarchical community occupancy model \citep{dorazio2005estimating, gelfand2005modelling}, species-specific parameters in both the ecological process model and observation models are treated as random effects arising from community level normal distributions with associated community level mean and variance parameters. For example, $\beta0_{i, t}$, the intercept on occurrence probabilities for species $i$ in year $t$, is modeled as 

\begin{equation}\label{comm-dist}
    \beta0_{i, t} \sim \text{Normal}({\mu}_{\beta0_t}, \sigma^2_{\beta0_t}),
\end{equation}

where ${\mu}_{\beta0_t}$ is the hyper-mean for occurrence probability (on the logit scale) of all species in the community in year $t$ (at average covariate values) and $\sigma^2_{\beta0_t}$ is the hyper-variance for occurrence probability across all species in the community in year $t$. Models for all other species-specific effects in the ecological and observation models are defined analogously. By treating species-specific effects as random, we improve estimates for both rare and abundant species \citep{zipkin2009impacts} while simultaneously estimating community level effects.

A further benefit of the hierarchical community modeling approach is the ability to easily calculate biodiversity metrics (e.g., alpha, beta diversity) from the latent occurrence state ($z_{i, j, t}$) that account for imperfect detection of species. Under a Bayesian framework, we can calculate any biodiversity metric as a derived parameter at each iteration of the MCMC to obtain a full posterior distribution from which we can obtain estimates with fully propagated uncertainty. For example, we can estimate species richness at each site $j$ and year $t$ at each iteration of the MCMC by summing the latent occurrence state ($z_{i, j, t}$) for all species. As a metric of beta diversity, we can calculate the Jaccard index \citep{magurran2013measuring}, which describes the similarity between two sites in terms of the number of species that occur at both sites. More specifically, we calculate the Jaccard index between site $j$ and $j'$ in year $t$ as 

\begin{equation}\label{jaccard}
    \text{JACCARD}_{j, j', t} = \frac{\sum_{i = 1}^Iz_{i, j, t} \cdot z_{i, j', t}}{\sum_{i = 1}^Iz_{i, j, t} + \sum_{i = 1}^Iz_{i, j', t} - \sum_{i = 1}^Iz_{i, j, t} \cdot z_{i, j', t}},
\end{equation}

which takes value 0 if the two sites have no species in common and value 1 if the same species occur at the two sites.

\subsubsection*{Data integration via joint likelihood}

We use a joint likelihood framework to integrate the replicated and nonreplicated detection-nondetection data sources into a single model (the ICOM; \citealt{miller2019recent}). To do this, we assume the likelihoods for the individual data sets are independent, conditional on the shared latent ecological process. This assumption can be interpreted as the detection of a species in one data set (conditional on the species being present) is independent of the detection of the species in any other data set \citep{schaub2011integrated, keryRoyle2020}. Thus, our full joint likelihood, conditional on the true, shared ecological process, is the product of the individual conditional likelihoods for each data set: 

\begin{equation}\label{jointLike}
    \text{L}_{\text{ICOM}}(\bm{\alpha0}, \bm{\alpha}, \bm{\gamma0}, \bm{\gamma} \mid \bm{z}, \bm{\beta0}, \bm{\beta}, \bm{y}, \bm{v}) = \text{L}_{\text{REP}}(\bm{\alpha0}, \bm{\alpha} \mid \bm{z}, \bm{\beta0}, \bm{\beta}, \bm{y}) \cdot \text{L}_{\text{NREP}}(\bm{\gamma0}, \bm{\gamma} \mid \bm{z}, \bm{\beta0}, \bm{\beta}, \bm{v}).
\end{equation}

\subsection*{Simulation study 1: Assessing benefits of integration}
 
We performed a simulation study to assess whether integration of multiple data sources in an ICOM framework could provide improved accuracy and precision for species and community level parameters compared to individual analyses under a range of realistic parameter values (Supplemental Information S1.1). We simulated data from one replicated data source with $K = 3$ replicates and two nonreplicated data sources, with the replicated data source having medium community-level detection probability (mean hyper-parameter = 0.5), one nonreplicated data source having low community-level detection probability (mean = 0.22), and the other having high community-level detection probability (mean = 0.78), which allowed us to compare the benefits of integration across varying qualities of nonreplicated data. We generated 100 replicates of each data source under the ICOM framework using a range of community-level ecological parameter values and subsequently drew simulated species data for $I = 25$ species for $T = 6$ years. We generated species' occurrence probabilities according to Equations \ref{psi1} and \ref{psit} with a single spatially-varying covariate. We generated detection processes for each data source as a function of a species and year specific intercept, and a species-specific effect of a spatio-temporally varying covariate unique to each data set. We simulated all covariates as normally distributed random variables with mean zero and standard deviation one. We generated each data source at 50 distinct sites that were randomly distributed across the range of covariates, resulting in a total of $J = 150$ sites. We compared model performance by fitting models individually for each of the seven unique combinations of the three data sources and subsequently computing the average bias (i.e., true simulated value minus estimated value) of the species-specific occurrence parameters across all 25 species and 100 simulations, as well as the average bias of the community level parameters. 

\subsection*{Simulation study 2: Assessing benefits of community modeling}
 
To evaluate the benefits of the hierarchical community model approach used in the ICOM, we assessed how species-level estimates from the ICOM compared to estimates from single species integrated distribution models (IDMs). The IDM took the same form as the ICOM except species-specific parameters were no longer random effects from a community level distribution, rather species-specific parameters were estimated individually in a model for each species. We simulated data from one replicated data source with $K = 3$ replicates and one nonreplicated data source, both with medium community-level detection probabilities (mean = 0.5). We simulated a community of $I = 25$ species, where each of the two data sources consisted of 50 unique locations sampled over $T = 6$ years, where the locations were randomly distributed across the range of a spatially varying covariate influencing occurrence. We assumed detection for both data sets was a function of a species and year specific intercept, and a species-specific spatio-temporally varying covariate unique to each data point. We generated species-specific occurrence and detection intercepts and covariate effects from uniform distributions (Supplemental Information S1.1), which allowed us to compare the ICOM to individual IDMs under the scenario when species level effects may not follow a normal distribution. We simulated 100 data sets from the community under realistic parameter values (Supplemental Information S1.1). We assessed model performance across the 100 simulated data sets by comparing the accuracy and precision of estimates from the ICOM to the IDMs. 
 
 \subsection*{Case study: foliage-gleaning birds in the White Mountains}
 
\begin{figure}[h]
     \centering
     \includegraphics[width = 15cm]{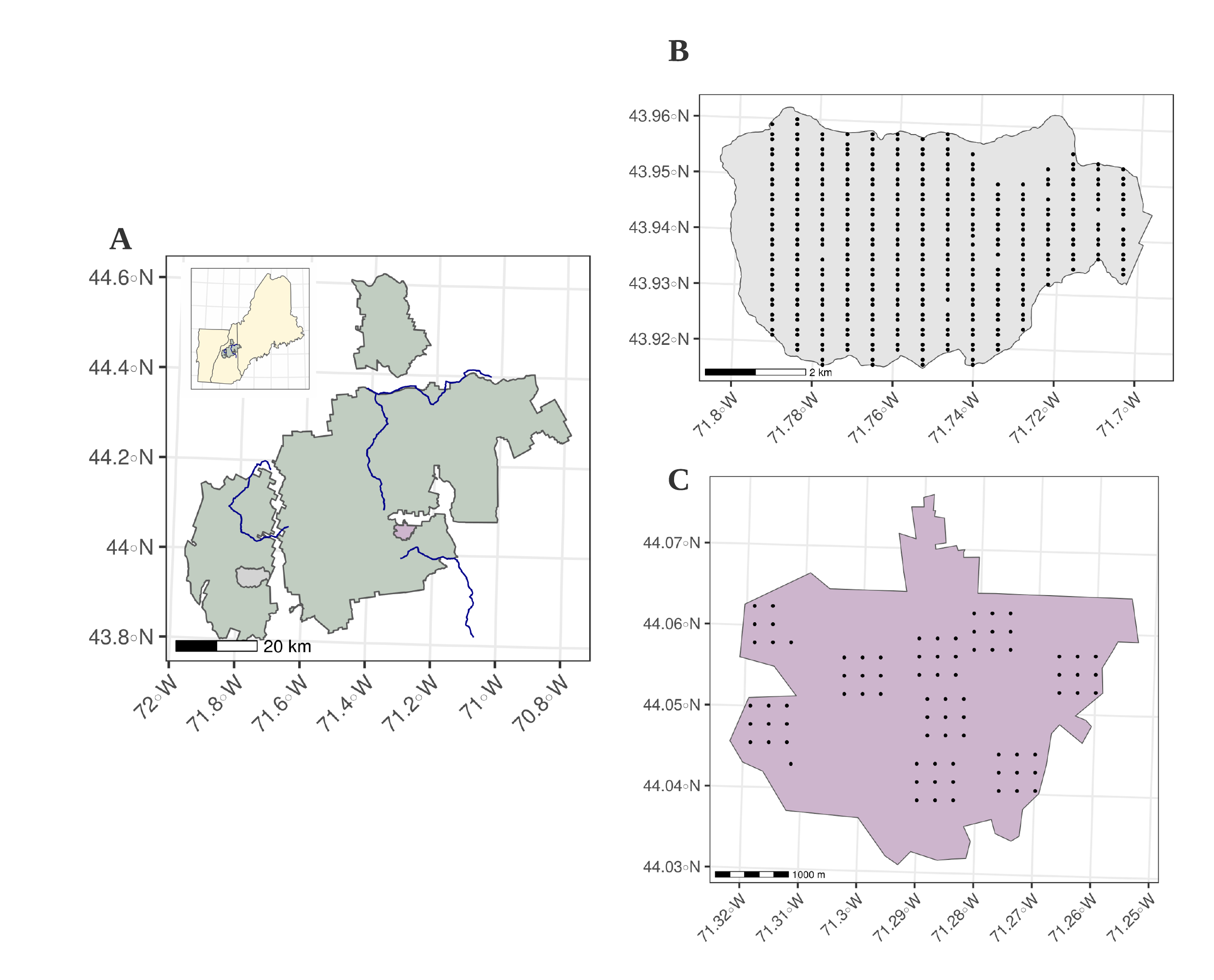}
     \caption{Study location for the case study. Panel (A) shows the White Mountain National Forest (shaded dark grey region) and the location of the Hubbard Brook Experimental Forest (HBEF; light grey region), the BBS routes (dark blue lines), and the NEON data from Bartlett Forest (purple region). Panel (B) shows the distribution of point count locations in HBEF, and Panel (C) shows the distribution of points in the NEON data set. Note different axis spacings across the three plots.}
     \label{fig:wmnf}
\end{figure}

We applied the ICOM to characterize temporal trends and spatial variability in individual species occurrence, species richness, and species composition of a community of twelve foliage-gleaning birds from 2010-2018 across the White Mountain National Forest using two replicated data sets and one nonreplicated data set (Figure \ref{fig:wmnf}). Our two replicated data sets come from the Hubbard Brook Experimental Forest (HBEF) and the National Ecological Observatory Network (NEON) at Bartlett Experimental Forest \citep{barnett2019terrestrial, neonData}, while our nonreplicated data set comes from the North American Breeding Bird Survey (BBS; \citealt{pardieck2020north}). At HBEF, observers performed three replicate surveys at each of 373 sites in each survey year to account for imperfect detection, while observers at NEON used a removal design at 81 sites to separate detection from occurrence. BBS observers performed point counts at 50 point count locations (called stops) along four routes (i.e., roads) that at least partially fell within the White Mountain National Forest, resulting in a total of 200 nonreplicated point count locations during each survey year. Integration of these three data sets is particularly valuable as each data source has clear advantages and disadvantages (Table \ref{tab:dataCharacteristics}) and cover disparate areas within the study region, and thus integration may yield parameter estimates more indicative of the entire White Mountains rather than analyses of the data sources independently. See Supplemental Information S1.2 for additional details on the three data sets. 

\begin{table}[ht] 
  \begin{center}
  \caption{Characteristics of the three data sources used to model occurrence dynamics of twelve foliage-gleaning birds in the White Mountain National Forest from 2010-2018. Forest Cover corresponds to the amount of forest within a 250m radius of a point count site. Values for elevation and forest cover are mean (minimum, maximum).}
  \label{tab:dataCharacteristics}
  \begin{tabular}{|c | c | c | c | }
    \toprule
    & HBEF & NEON & BBS \\
    \midrule
    Data type & Replicated & Replicated & Nonreplicated \\
    Years & 2010-2018 & 2015-2018 & 2010-2018 \\
    Number of sites & 373 & 81 & 200 \\
    Elevation (m) & 607 (240, 932) & 432 (268, 766) & 352 (134, 917)\\
    Forest Cover (\%) & 97.7 (71, 100) & 94.2 (75, 100) & 70.6 (0, 92) \\
    Survey Location & Experimental forest & Experimental forest & Roadside \\
    \bottomrule
  \end{tabular}
  \end{center}
\end{table}

We modeled occurrence dynamics for the following twelve foliage-gleaning bird species: American Redstart (\textit{Setophaga ruticilla}), Black-and-white Warbler (\textit{Mniotilta varia}), Blue-headed Vireo (\textit{Vireo solitarius}), Blackburnian Warbler (\textit{Setophaga fusca}), Blackpoll Warbler (\textit{Setophaga striata}), Black-throated Blue Warbler (\textit{Setophaga caerulescens}), Black-throated Green Warbler (\textit{Setophaga virens}), Canada Warbler (\textit{Cardellina canadensis}), Magnolia Warbler (\textit{Setophaga magnolia}), Nashville Warbler (\textit{Leiothlypis ruficapilla}), Ovenbird (\textit{Seiurus aurocapilla}), and Red-eyed Vireo (\textit{Vireo olivaceus}). We specified species occurrence following Equation \ref{initOcc}, with occurrence probability in the first year, $\psi_{i, j, 1}$, modeled according to 

\begin{equation}
   \text{logit}(\psi_{i, j, 1}) = \beta0_{i, 1} + \beta1_{i} \cdot \text{ELEV}_j + \beta2_i \cdot \text{ELEV}^2_j + \beta3_i \cdot \text{FOR}_j, 
\end{equation}

where $\beta0_{i, 1}$ is the species-specific intercept in year 1, and $\beta1_i$, $\beta2_i$, and $\beta3_i$ are species-specific effects of elevation (ELEV, linear and quadratic) and local forest cover within a 250m radius (FOR), respectively. Occurrence in subsequent years is modeled analogously with a year-specific intercept and an autologistic parameter following Equation \ref{psit}. We extracted elevation data at a 30 $\times$ 30 m resolution from the National Elevation Dataset \citep{gesch2002national} and associated each point count site with the elevation at the center of the point count. We used the National Land Cover Database \citep{Homer2015} to determine the amount of local forest cover in 2016 within a 250m radius of each point count location. To compute species-specific temporal trend estimates, we performed a post-hoc linear regression using the average occurrence probability of each species during each year as a response variable and year as a covariate. Under a Bayesian framework, we obtain full uncertainty propagation by calculating the trend for each posterior sample of the average occurrence probabilities (Supplemental Information S1.3). All species-specific occurrence intercepts and regression coefficients were modeled hierarchically following Equation \ref{comm-dist}. 
 
We incorporated multiple covariates in the conditional likelihoods of each data type to account for variation in detection rates following the species' detection models described in the ``Integrated community occupancy model'' section. For the HBEF and NEON data, we included species and year-specific intercepts, a species-specific linear effect of the time of the survey, and species-specific linear and quadratic effects of the day of the survey. For the BBS data, we modeled detection as a function of a species and year specific intercept, species-specific linear and quadratic effects of day of survey, and a random observer effect to account for variation in detection among observers. All detection covariates were modeled hierarchically following Equation \ref{comm-dist}. Species and year specific intercepts were also modeled hierarchically, but were drawn from a single distribution for all species and years within each data set (Supplemental Information S3).  

\subsubsection*{Goodness of fit and model validation}

We assessed model fit for the case study using a Bayesian p-value approach with a Chi-square fit statistic (Supplemental Information S4). We used two-fold cross validation with the log predictive density \citep{vehtari2017practical} as a predictive performance metric to assess the out-of-sample predictive performance of the full ICOM compared to six models using subsets of the three data sets for the case study. Assessing out-of-sample predictive performance with occupancy models presents additional complexities since the ecological state of interest is not directly observed \citep{zipkin2012evaluating}. For a given data set, we compared the occurrence predictions at the hold out locations to the occurrence values generated from models that were fit using the data at the hold out locations. To account for model uncertainty, we compared the occurrence predictions individually to latent occurrence values generated from the subset of the seven models that used the data set in the model fitting process (see Supplemental Information S5 for details). We summarize predictive performance for each species and the entire community individually at each data set location, as well as across the entire study region (i.e., White Mountains). We used a similar approach to compare the performance of the ICOM to individual species IDMs (Supplemental Information S5). 

\subsection*{Model implementation}

We estimated the parameters in all model versions (simulations and case study) with a Bayesian framework using Markov Chain Monte Carlo (MCMC). We fit the models in NIMBLE \citep{de2017Programming, nimble} within the \texttt{R} statistical environment \citep{rSoftware} using vague priors for all hyper-parameters (Supplemental Information S1.3). For all simulations, we ran three chains, each with 20,000 iterations with a burn-in period of 10,000 iterations and a thinning rate of four. For the case study, we ran models for three chains of 450,000 iterations with a burn-in period of 200,000 iterations and a thinning rate of 20, resulting in a total of 52,000 samples from the posterior distribution. We assessed model convergence using the Gelman-Rubin R-hat diagnostic \citep{brooks1998} and visual assessment of trace plots using the \verb+coda+ package \citep{coda}. 

\section*{Results}
 
\subsection*{Simulations}
 
\begin{figure}[h]
     \centering
     \includegraphics[width = 15cm]{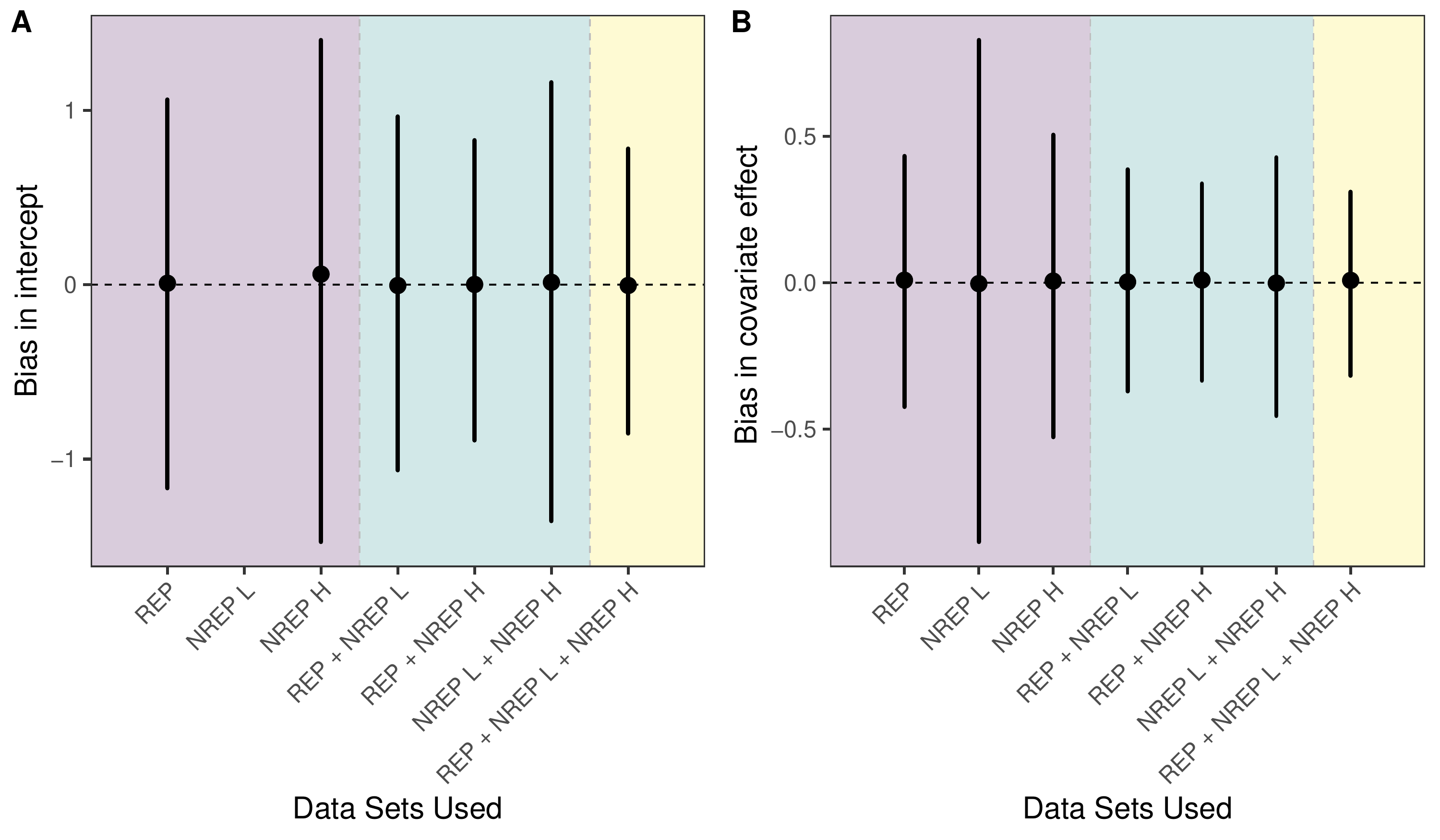}
     \caption{Sampling distribution of estimated bias in simulated species level occurrence intercepts (A) and covariate effects (B) under models using different combinations of a replicated (REP) data set and nonreplicated data sets with low (NREP L) and high (NREP H) detection probability. Points represent the median bias (posterior mean - true simulated value) in a species-level effect across 100 simulations for a community of 25 species. Lines represent the 95\% quantiles of the bias values. The intercept parameter using only NREP L is not shown as it failed to converge.}
     \label{fig:sim-1-sp-fig}
\end{figure}

The ICOM using one replicated data set, one nonreplicated data set with low average detection probability across the community, and one nonreplicated data set with high detection probability yielded unbiased estimates and was generally more precise in community and species-level occurrence parameter estimates than models using smaller combinations of the three data sets or data sets individually (Figure \ref{fig:sim-1-sp-fig}, Supplemental Figure S1). Patterns were similar across community effects and species-specific effects, with increases in precision more prominent in species-level effects. Despite the general improvement in estimates found when integrating all three data sources, models using only two data sources, particularly the replicated data source and the nonreplicated data source with high detection probability, also yielded estimates with low bias and high precision. Models using only the nonreplicated data source with low detection probability often failed to converge, while models using only the nonreplicated data source with high detection probability mostly converged but were less precise than models using replicated data (and with essentially unidentifiable intercept values; Figure \ref{fig:sim-1-sp-fig}A). 

The ICOM also led to substantial improvements in precision of parameter estimates compared to single species IDMs (Table \ref{tab:simIDM}, Supplemental Figure S2). Species-level IDMs provided slightly more accurate estimates of species-specific occurrence parameters for some species as compared to the ICOM, which is a result of Bayesian shrinkage (i.e., borrowing strength) driving species-level parameters closer to the community average for those species with extreme parameter values in the ICOM. However, the true species-level parameters were contained within the 95\% Bayesian credible interval of the estimated parameter values across 94.9\% of all simulations, indicating that this loss in accuracy is negligible. Further, we simulated species-level effects from a uniform distribution rather than a normal distribution, which led to more extreme species values. Losses in accuracy would be much lower for communities of species where the normal assumption is adequate. Thus, in addition to providing community-level parameter effects, the ICOM provides more precise estimates of species-specific effects compared to IDMs with only minor losses in accuracy for extreme species.

\begin{table}[ht] 
  \begin{center}
  \caption{Precision and accuracy of species-specific parameter estimates when using the integrated community occupancy model (ICOM) compared to a single species integrated distribution model (IDM) for a simulated community of 25 species over six years across 100 simulations with one replicated (REP) data set and one nonreplicated (NREP) data set. Precision improvement is the percentage improvement in precision when using the ICOM compared to the IDM, where precision is defined as the difference between the 2.5\% and 97.5\% quantiles of the posterior means. Bias is the average magnitude of the posterior means minus the true simulated value. Values are averaged across all 25 species and six years.}
  \label{tab:simIDM}
  \begin{tabular}{|c | c | c | c | c |}
    \toprule
    Parameter & Precision & ICOM Bias & IDM Bias & Parameter \\
    & Improvement (\%) & & & \\
    \midrule
    $\gamma0_{i}$ & 41.9 & 0.235 & 0.101 & NREP detection intercept \\
    $\gamma1_{i}$ & 33.4 & 0.070 & 0.027 & NREP detection covariate \\
    $\phi_i$ & 30.0 & 0.173 & 0.121 & Auto-logistic \\
    $\beta0_i$ & 29.2 & 0.173 & 0.108 & Occurrence intercept \\
    $\beta1_i$ & 22.5 & 0.042 & 0.017 & Occurrence covariate \\
    $\alpha0_i$ & 18.8 & 0.104 & 0.044 & REP detection intercept \\
    $\alpha1_i$ & 9.63 & 0.023 & 0.012 & REP detection covariate \\
    \bottomrule
  \end{tabular}
  \end{center}
\end{table}

\subsection*{Case Study}

\begin{figure}
    \centering
    \includegraphics[width = 15cm]{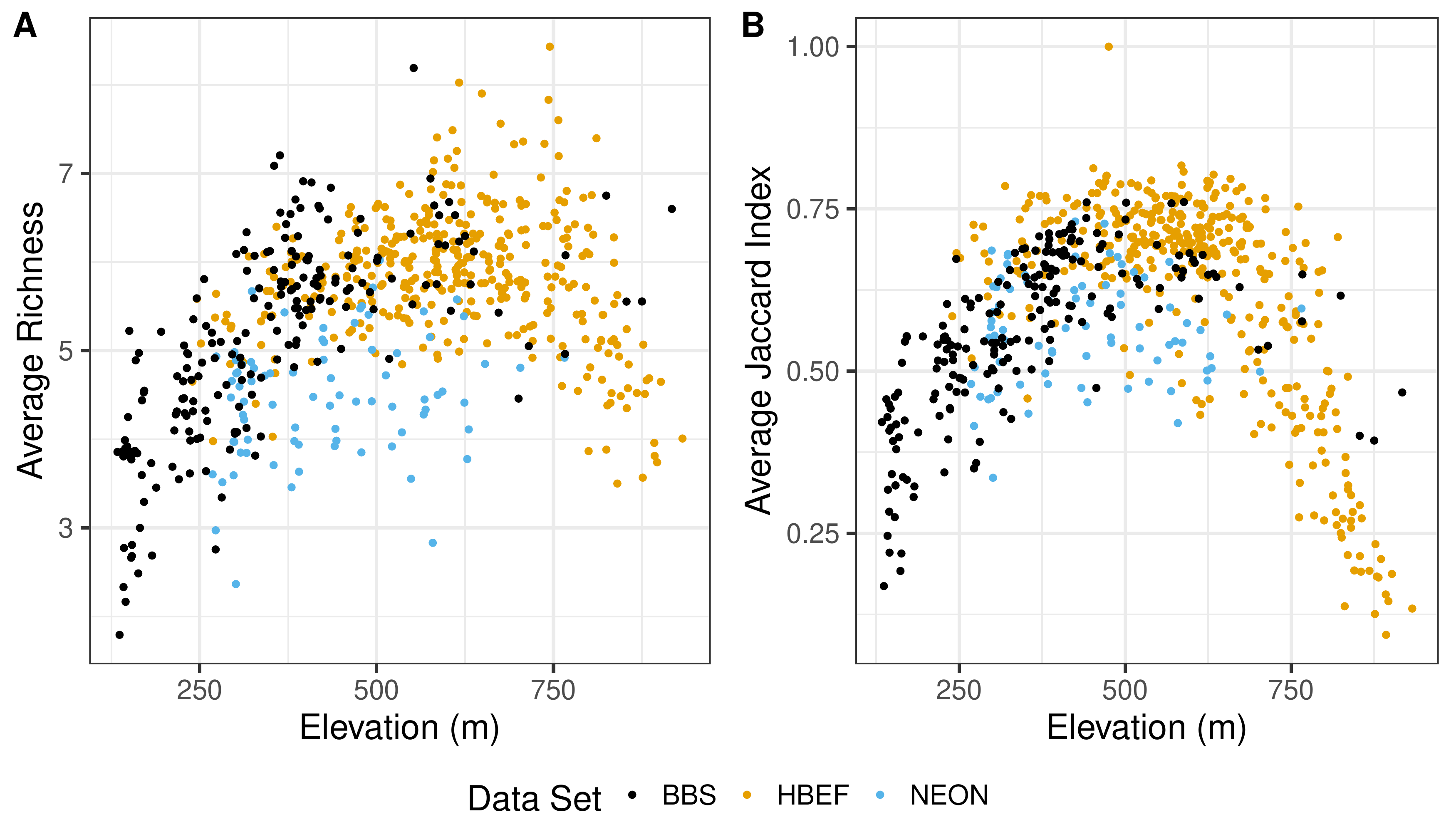}
    \caption{Estimated average site-level species richness (A) and Jaccard index (B) of a community of twelve foliage-gleaning bird species in the Hubbard Brook Experimental Forest (HBEF). Points are posterior means. Jaccard index values are relative to a single site in HBEF with value 1, with 0 indicating no species in common to the reference site, and 1 indicating identical community composition to the reference site.}
    \label{fig:hbef-richness}
\end{figure}

\begin{figure}
    \centering
    \includegraphics[width = 15cm]{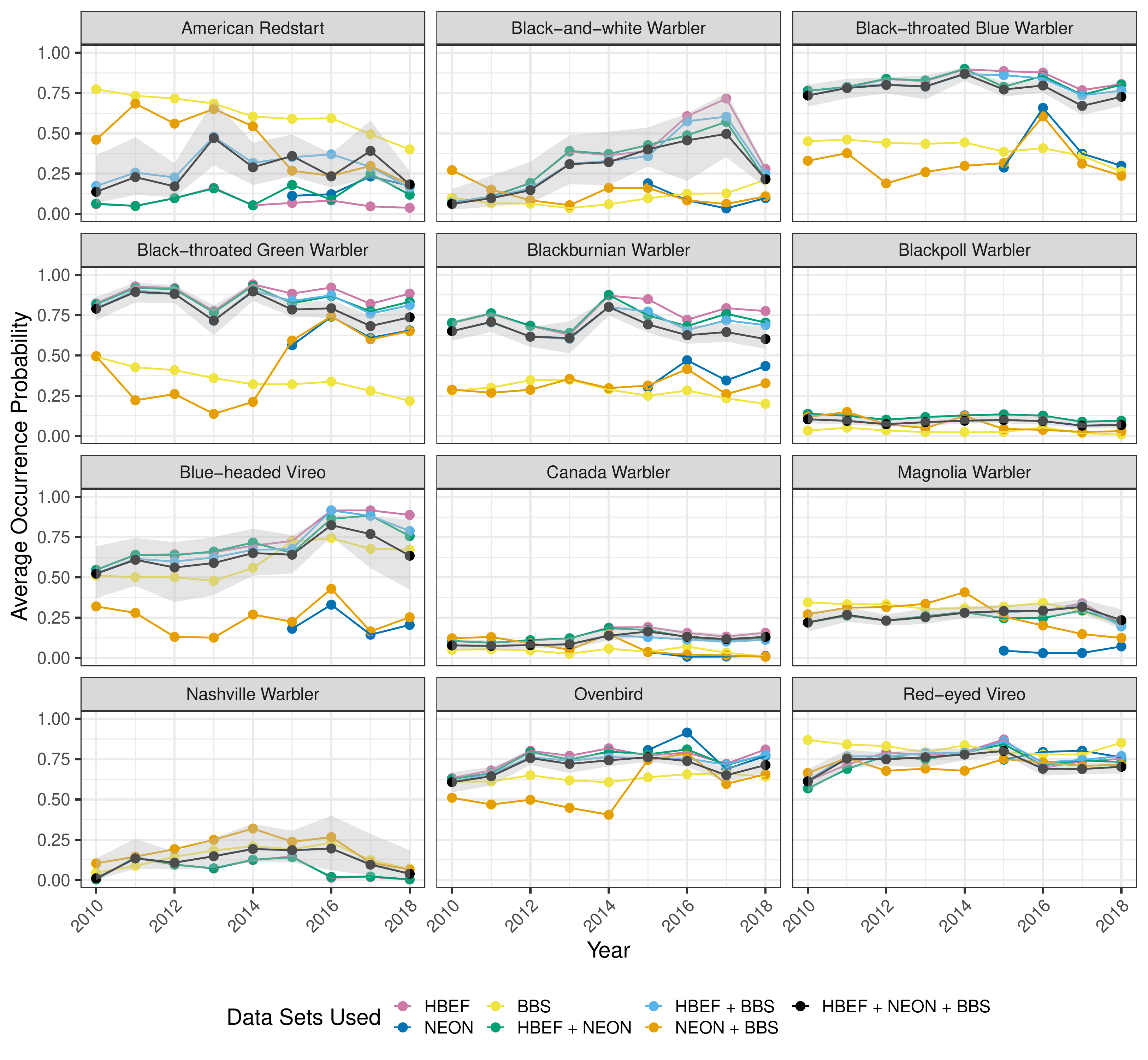}
    \caption{Average occurrence probabilities of twelve foliage-gleaning bird species in the White Mountain National Forest from 2010 to 2018 from models using different subsets of the three data sources. Points show posterior mean occurrence probabilities averaged across all sites in a given year. Gray shaded region indicates the 95\% credible interval for the model with all three data sets.}
    \label{fig:wmnf-trend-comp}
\end{figure}

The ICOM estimated variable trends in occurrence for the twelve foliage-gleaning bird species across the White Mountain National Forest, with five species having >75\% probability of increasing occurrence rates and three species having >75\% probability of decreasing (Supplemental Figure S3) from 2010-2018. Community level parameters revealed that average occurrence probability peaked at medium elevations and higher amounts of local forest cover across the community, although species-specific parameters were highly variable (Supplemental Table S3). Occurrence probabilities peaked at a variety of elevations across the twelve species (Supplemental Figure S4), which resulted in species richness being maximized at medium elevations (600-800m; Figure \ref{fig:hbef-richness}A). Species composition of the community, as measured by the Jaccard index, largely followed similar patterns (Figure \ref{fig:hbef-richness}B). Estimated trends in species-specific occurrence probabilities were highly dependent on the data sets included in the model (Figure \ref{fig:wmnf-trend-comp}), suggesting important spatial variability across the White Mountains. For example, while Red-eyed Vireo occurrence showed consistent trends across models from all data source combinations, trends for the Black-throated Blue Warbler and Black-throated Green Warbler were stable in estimates from most data combinations, but occurrence probabilities were estimated to have declined over this time period in a model using only BBS data.  

Integration of all data sources in the ICOM yielded better predictive performance for the community of birds across the three data set locations than models using only a subset of the available data (Table \ref{tab:crossVal}). This is likely a result of both a larger number of detections and a wider range of the covariate space when using all three data sources. The model using only HBEF and BBS data had the highest predictive performance for data at HBEF, while the model using NEON data only had the highest predictive performance for data at NEON. The ICOM had higher predictive performance compared to single species IDMs across all 12 species for the three data sets, and outperformed single species IDMs individually for each species for 64\%, 91\%, and 100\% of the species at HBEF, BBS, and NEON, respectively (Supplemental Table S6). 

\begin{table}[ht] 
  \begin{center}
  \caption{Two-fold cross validation results comparing predictive performance across models using different combinations of the three data sets. The fitted model is shown in the first column and log predictive density measures are shown for the entire community with each data set and across all data sets. Values in parentheses show the average rank of the model for an individual species across all models, with 1 indicating the model is the best for all species and 7 indicating the the model is worst for all species. Bold values indicate the best performing model for each individual data set. Predictive performance of the model using only NEON data is only assessed at NEON locations because NEON data are only available for four of the nine study years.}
  \label{tab:crossVal}
  \begin{tabular}{c  c  c  c | c}
    \toprule
    Model & HBEF & BBS & NEON & All \\
    \midrule
    HBEF & -10174 (3.75) & -6200 (4.67) & -947 (4.75) & -17321 (4.5) \\
    NEON & - & - & \textbf{-792} (\textbf{2.83}) & - \\
    BBS & -13876 (3.75) & -5702 (3.5) & -1148 (5.08)  & -20726 (3.83) \\
    HBEF+NEON & -10097 (3.58) & -5878 (3.33) & -852 (3.67) & -16829 (3.33) \\
    HBEF+BBS & \textbf{-9732} (\textbf{2.67}) & -5717 (3.67) & -934 (4.42) & -16383 (2.67) \\
    NEON+BBS & -12560 (4.5) & -5759 (3.25) & -801 (3.83) & -19121 (4.33) \\
    HBEF+NEON+BBS & -9767 (2.75) & \textbf{-5691} (\textbf{2.58}) & -836 (3.42) & \textbf{-16294} (\textbf{2.33}) \\
    \bottomrule
  \end{tabular}
  \end{center}
\end{table}

\section*{Discussion}

Understanding species distributions and occurrence dynamics of multiple species in a community is an important task for biodiversity conservation \citep{guillera2019inferring}. Monitoring programs collect different types of data that vary in amount, spatial extent, quality, and information content, and incorporating these varied data into a unified analysis can yield improved estimates on quantities of interest \citep{zipkin2018synthesizing}. We developed an ``integrated community occupancy model'' (ICOM) that uses replicated and nonreplicated detection-nondetection data to simultaneously provide inferences on species-specific and community dynamics. Using simulations and empirical bird data, we showed that the ICOM can provide more accurate and precise estimates of occurrence dynamics than analyses using single data sources (Figure \ref{fig:sim-1-sp-fig}) or single species models (Table \ref{tab:simIDM}) as well as improved predictive performance across a region of interest (Table \ref{tab:crossVal}, Supplemental Table S6).

In our simulation study, the ICOM using one replicated data set, one nonreplicated data set with low detection probability, and one replicated data set with high detection probability provided unbiased and generally more precise parameter estimates than models using a subset of the three data sources (Figure \ref{fig:sim-1-sp-fig}), which aligns with previous single species data integration work \citep{fletcher2019practical}. Despite this general improvement in the simulation study, integrating the single replicated data source with the high detection nonreplicated data source yielded comparable accuracy and precision to the model using one replicated and two nonreplicated data sources. This suggests that integrating a data source of particularly low quality (e.g., nonreplicated, potentially large detection variability) with higher quality data sources may not yield any practical benefits \citep{simmonds2020more}. We generated each data source at random locations across the range of the covariate with no systematic bias in sampling locations. In reality, many sources of detection-nondetection data are spatially biased (e.g., collected along road transects or near locations with high human density), which can lead to a narrow range of habitat variables (e.g., forest cover) that drive species and community occurrence dynamics. Such biases can result in different species being observed across data sets or estimated parameters that are either biased, only span a portion of the covariate range of interest, or are not indicative of the larger spatial areas of interest \citep{conn2017confronting}. By integrating disparate data sources from multiple locations in the ICOM, we can increase the likelihood that the sampled sites vary along important ecological and environmental gradients, which in turn enables more precise and accurate inference on the environmental conditions driving species occurrence.

In the foliage-gleaning bird case study, we used a two-fold cross validation approach to show that integration of all three data sets yielded the best overall predictive performance across the White Mountain National Forest compared to models using smaller subsets of the three data sources (Table \ref{tab:crossVal}). Further, the ICOM generally yielded improved predictive performance for individual species and the overall community compared to single species integrated distribution models (Supplemental Table S5). In contrast to the simulation study, parameter precision was not always highest when integrating all three data sets in the case study (Supplemental Tables S4, S5). Additionally, models with smaller subsets of the three data sources outperformed the ICOM with all three data sets individually for the HBEF and NEON data sets, although the improvements in predictive performance were not large (Table \ref{tab:crossVal}). This is likely a result of different covariate ranges among the three data sources and only having NEON data for a subset of the time period of interest. For example, precision of species-specific effects of local forest cover was highest for the model using HBEF and NEON data. Sites at HBEF and NEON have low variability in forest cover, with average forest cover of sites being 94.2\% and 97.7\%, respectively. However, variability in forest cover is higher at BBS sites, which likely explains why precision is lower when incorporating BBS data in the model. For most intercept parameters at the community and species-specific level, models using either NEON data alone or NEON and BBS were often the most precise. For these models, information for separating detection probability from true occurrence comes solely from NEON data, which are only available for four of the nine years of the study period. This smaller time period likely results in less unexplained variability in occurrence probability and detection across years, which in turn leads to more precise intercept estimates. While the smaller covariate and temporal ranges marginally increased precision, the ICOM using all three data sets enables inference across the entire temporal period of interest as well as across a broader range of covariates, which makes estimates on trends and spatial patterns across the White Mountains more informative. 

The large variation in temporal trends when using different combinations of the three data sources suggests spatially-varying occurrence trends for the community of twelve foliage-gleaning birds. NEON data were only available from 2015-2018, which may account for differences in estimated trends from the model using only NEON data compared to other models. BBS data were sampled along road transects and thus have less forest cover than both the NEON and HBEF sites, suggesting that occurrence trends could vary as a result of differences in amount of local forest cover and/or proximity to roads \citep{furnas2020rapid}. Estimates from the full ICOM are a weighted average across heterogeneity in trends across the region, where the weights are determined by the amount of the different data sources \citep{fletcher2019practical}. In our case study, the HBEF data source comprised a majority of observations (77\%; Supplemental Table S2) and thus contributed the most to estimates of model parameters, which we deemed acceptable because the HBEF data are a high-quality replicated data source. Alternatively, a profiling approach could be used within the MCMC sampler to change the weights for each data set similar to the maximum likelihood approach of \citealt{fletcher2019practical}. By including information from multiple data sources within a region, the ICOM yields area-wide averaged species-specific trends. If an area-wide averaged trend is not desired, trends from different spatial locations could be estimated hierarchically in a multi-region framework \citep{doser2021trends} or treated as spatially varying coefficients to explicitly model the spatial heterogeneity \citep{finley2011comparing}. 

We envision numerous methodological extensions and ecological applications of the ICOM framework. While our case study used three data sources arising from the same method (i.e., point count surveys), the ICOM can incorporate detection-nondetection data from different data collection approaches (e.g., camera traps, autonomous recording units, citizen science checklists) in an analogous manner. Similar integrated models could be developed to estimate alternative ecological processes such as abundance of multiple species within a community by extending single species integrated models that use distance sampling \citep{farr2021integrating}, acoustic recordings \citep{doser2021Integrating}, or capture-recapture data \citep{chandler2014spatially}. If regional species pools differ between data source locations, the ICOM could be adapted to a multiregion framework \citep{Sutherland2016}. Given our model's ability to estimate species and community level effects, the ICOM can be applied to help elucidate individual species sensitivities to various global change drivers, determine factors causing shifts in taxonomic or functional diversity, or forecast future species and community shifts under varying climate and land use change scenarios to help prioritize conservation strategies. In particular, the ICOM will assist multi-species conservation planning by providing estimates for rare species that lack adequate data for common analysis approaches, while simultaneously obtaining inference on an entire community that may elucidate specific species traits linked to occurrence trends and be more indicative of large-scale biodiversity change.

The benefits of data integration for multi-species detection-nondetection data sets will depend on characteristics and goals of each specific study. While the ICOM generally leads to improved inference for species and community level effects, integrating multiple data sets leads to increased computation times and potential  difficulties in model convergence. When determining what data sources to use in an ICOM, we recommend considering the following factors: (1) the amount of the different data sources within the area of interest and how they are distributed across the range of ecological and environmental gradients; (2) the precision of estimates required for the analysis objectives; (3) amount of time and computing power available to run the models; (4) the spatial resolution of each data source; and (5) the quality and information content) of each data source (e.g., replicated vs nonreplicated, large detection variability vs. standardized protocol). For example, if high quality replicated data exist across an adequate number of years and spatial locations that are distributed across potential habitat, a community model using only these data may suffice to accomplish research objectives. Simulation studies based on the specific data sets and sample sizes for a given research study can help to determine the ideal combinations of data required for a given study objective. 

As changes in environmental and climate conditions continue globally, continued development of monitoring and analysis techniques that can effectively produce accurate and precise estimates of biodiversity metrics are needed to understand global change impacts and develop appropriate mitigation plans. Our integrated community occupancy models provide a new approach to simultaneously analyze multi-species data from numerous available sources. This framework can be used to elucidate both species specific and community level dynamics, improving understanding of the mechanisms driving biodiversity shifts and informing appropriate management and conservation actions to address global change.

\section*{Authors' Contributions}
 
JWD and EFZ developed the modeling framework with critical insight provided by WL. TSS and MTH assisted in data management and preparation. JWD performed all analyses and led writing of the manuscript. All authors contributed critically to the drafts and gave final approval for publication.

\section*{Acknowledgements}

We declare no conflicts of interest. This manuscript is a contribution of the Hubbard Brook Ecosystem Study. Hubbard Brook is part of the Long- Term Ecological Research network, which is supported by the U.S. National Science Foundation. The Hubbard Brook Experimental Forest is operated and maintained by the USDA Forest Service, Northern Research Station, Newtown Square, PA. This work was supported by National Science Foundation grant DBI-1954406.

\section*{Data Availability}

All data and code associated with this manuscript are posted on GitHub (\url{https://github.com/zipkinlab/Doser_etal_2021_InReview}) and will be archived on Zenodo upon acceptance. 

\bibliographystyle{apalike}
\bibliography{references}

\begin{thebibliography}{}

\bibitem[Barnett et~al., 2019]{barnett2019terrestrial}
Barnett, D.~T., Duffy, P.~A., Schimel, D.~S., Krauss, R.~E., Irvine, K.~M.,
  Davis, F.~W., Gross, J.~E., Azuaje, E.~I., Thorpe, A.~S., Gudex-Cross, D.,
  et~al. (2019).
\newblock The terrestrial organism and biogeochemistry spatial sampling design
  for the national ecological observatory network.
\newblock {\em Ecosphere}, 10(2):e02540.

\bibitem[Barraquand and Gimenez, 2019]{barraquand2019integrating}
Barraquand, F. and Gimenez, O. (2019).
\newblock Integrating multiple data sources to fit matrix population models for
  interacting species.
\newblock {\em Ecological modelling}, 411:108713.

\bibitem[Brooks and Gelman, 1998]{brooks1998}
Brooks, S.~P. and Gelman, A. (1998).
\newblock General methods for monitoring convergence of iterative simulations.
\newblock {\em Journal of Computational and Graphical Statistics},
  7(4):434--455.

\bibitem[Chandler and Clark, 2014]{chandler2014spatially}
Chandler, R.~B. and Clark, J.~D. (2014).
\newblock Spatially explicit integrated population models.
\newblock {\em Methods in Ecology and Evolution}, 5(12):1351--1360.

\bibitem[Clark et~al., 2017]{clark2017generalized}
Clark, J.~S., Nemergut, D., Seyednasrollah, B., Turner, P.~J., and Zhang, S.
  (2017).
\newblock Generalized joint attribute modeling for biodiversity analysis:
  Median-zero, multivariate, multifarious data.
\newblock {\em Ecological Monographs}, 87(1):34--56.

\bibitem[Clark, 2021]{clark2021large}
Clark, T.~J. (2021).
\newblock {\em {Large carnivore recolonization reshapes population and
  community dynamics in the Rocky Mountains: Implications for harvest
  management}}.
\newblock PhD thesis, University of Montana.

\bibitem[Conn et~al., 2017]{conn2017confronting}
Conn, P.~B., Thorson, J.~T., and Johnson, D.~S. (2017).
\newblock Confronting preferential sampling when analysing population
  distributions: diagnosis and model-based triage.
\newblock {\em Methods in Ecology and Evolution}, 8(11):1535--1546.

\bibitem[{de Valpine} et~al., 2021]{nimble}
{de Valpine}, P., Paciorek, C., Turek, D., Michaud, N., Anderson-Bergman, C.,
  Obermeyer, F., {Wehrhahn Cortes}, C., Rodrìguez, A., {Temple Lang}, D., and
  Paganin, S. (2021).
\newblock {\em {NIMBLE}: {MCMC}, Particle Filtering, and Programmable
  Hierarchical Modeling}.
\newblock {R} package version 0.11.1.

\bibitem[{de Valpine} et~al., 2017]{de2017Programming}
{de Valpine}, P., Turek, D., Paciorek, C., Anderson-Bergman, C., {Temple Lang},
  D., and Bodik, R. (2017).
\newblock Programming with models: writing statistical algorithms for general
  model structures with {NIMBLE}.
\newblock {\em Journal of Computational and Graphical Statistics}, 26:403--413.

\bibitem[Dorazio, 2014]{dorazio2014accounting}
Dorazio, R.~M. (2014).
\newblock Accounting for imperfect detection and survey bias in statistical
  analysis of presence-only data.
\newblock {\em Global Ecology and Biogeography}, 23(12):1472--1484.

\bibitem[Dorazio et~al., 2010]{dorazio2010models}
Dorazio, R.~M., Kery, M., Royle, J.~A., and Plattner, M. (2010).
\newblock Models for inference in dynamic metacommunity systems.
\newblock {\em Ecology}, 91(8):2466--2475.

\bibitem[Dorazio and Royle, 2005]{dorazio2005estimating}
Dorazio, R.~M. and Royle, J.~A. (2005).
\newblock Estimating size and composition of biological communities by modeling
  the occurrence of species.
\newblock {\em Journal of the American Statistical Association},
  100(470):389--398.

\bibitem[Doser et~al., 2021a]{doser2021Integrating}
Doser, J.~W., Finley, A.~O., Weed, A.~S., and Zipkin, E.~F. (2021a).
\newblock Integrating automated acoustic vocalization data and point count
  surveys for estimation of bird abundance.
\newblock {\em Methods in Ecology and Evolution}.

\bibitem[Doser et~al., 2021b]{doser2021trends}
Doser, J.~W., Weed, A.~S., Zipkin, E.~F., Miller, K.~M., and Finley, A.~O.
  (2021b).
\newblock Trends in bird abundance differ among protected forests but not bird
  guilds.
\newblock {\em Ecological Applications}, page e2377.

\bibitem[Farr et~al., 2021]{farr2021integrating}
Farr, M.~T., Green, D.~S., Holekamp, K.~E., and Zipkin, E.~F. (2021).
\newblock Integrating distance sampling and presence-only data to estimate
  species abundance.
\newblock {\em Ecology}, 102(1):e03204.

\bibitem[Finley, 2011]{finley2011comparing}
Finley, A.~O. (2011).
\newblock Comparing spatially-varying coefficients models for analysis of
  ecological data with non-stationary and anisotropic residual dependence.
\newblock {\em Methods in Ecology and Evolution}, 2(2):143--154.

\bibitem[Fletcher~Jr et~al., 2019]{fletcher2019practical}
Fletcher~Jr, R.~J., Hefley, T.~J., Robertson, E.~P., Zuckerberg, B., McCleery,
  R.~A., and Dorazio, R.~M. (2019).
\newblock A practical guide for combining data to model species distributions.
\newblock {\em Ecology}, 100(6):e02710.

\bibitem[Furnas, 2020]{furnas2020rapid}
Furnas, B.~J. (2020).
\newblock Rapid and varied responses of songbirds to climate change in
  california coniferous forests.
\newblock {\em Biological Conservation}, 241:108347.

\bibitem[Gallo et~al., 2017]{gallo2017mammal}
Gallo, T., Fidino, M., Lehrer, E.~W., and Magle, S.~B. (2017).
\newblock Mammal diversity and metacommunity dynamics in urban green spaces:
  implications for urban wildlife conservation.
\newblock {\em Ecological Applications}, 27(8):2330--2341.

\bibitem[Gelfand et~al., 2005]{gelfand2005modelling}
Gelfand, A.~E., Schmidt, A.~M., Wu, S., Silander~Jr, J.~A., Latimer, A., and
  Rebelo, A.~G. (2005).
\newblock Modelling species diversity through species level hierarchical
  modelling.
\newblock {\em Journal of the Royal Statistical Society: Series C (Applied
  Statistics)}, 54(1):1--20.

\bibitem[Gesch et~al., 2002]{gesch2002national}
Gesch, D., Oimoen, M., Greenlee, S., Nelson, C., Steuck, M., and Tyler, D.
  (2002).
\newblock The national elevation dataset.
\newblock {\em Photogrammetric engineering and remote sensing}, 68(1):5--32.

\bibitem[Guillera-Arroita et~al., 2019]{guillera2019inferring}
Guillera-Arroita, G., K{\'e}ry, M., and Lahoz-Monfort, J.~J. (2019).
\newblock Inferring species richness using multispecies occupancy modeling:
  Estimation performance and interpretation.
\newblock {\em Ecology and evolution}, 9(2):780--792.

\bibitem[Homer et~al., 2015]{Homer2015}
Homer, C., Dewitz, J., Yang, L., Jin, S., Danielson, P., Xian, G., Coulston,
  J., Herold, N., Wickham, J., and Megown, K. (2015).
\newblock Completion of the 2011 national land cover database for the
  conterminous united states--representing a decade of land cover change
  information.
\newblock {\em Photogrammetric Engineering \& Remote Sensing}, 81(5):345--354.

\bibitem[K{\'e}ry and Royle, 2020]{keryRoyle2020}
K{\'e}ry, M. and Royle, J.~A. (2020).
\newblock {\em Applied hierarchical modeling in ecology: Analysis of
  distribution, abundance, and species richness in R and BUGS: Volume 2:
  Dynamic and advanced models.}
\newblock Academic Press.

\bibitem[Lahoz-Monfort et~al., 2017]{lahoz2017bringing}
Lahoz-Monfort, J.~J., Harris, M.~P., Wanless, S., Freeman, S.~N., and Morgan,
  B.~J. (2017).
\newblock Bringing it all together: multi-species integrated population
  modelling of a breeding community.
\newblock {\em Journal of Agricultural, Biological and Environmental
  Statistics}, 22(2):140--160.

\bibitem[Lele et~al., 2012]{lele2012dealing}
Lele, S.~R., Moreno, M., and Bayne, E. (2012).
\newblock Dealing with detection error in site occupancy surveys: what can we
  do with a single survey?
\newblock {\em Journal of Plant Ecology}, 5(1):22--31.

\bibitem[MacKenzie et~al., 2002]{mackenzie2002estimating}
MacKenzie, D.~I., Nichols, J.~D., Lachman, G.~B., Droege, S., Andrew~Royle, J.,
  and Langtimm, C.~A. (2002).
\newblock Estimating site occupancy rates when detection probabilities are less
  than one.
\newblock {\em Ecology}, 83(8):2248--2255.

\bibitem[MacKenzie and Royle, 2005]{mackenzie2005designing}
MacKenzie, D.~I. and Royle, J.~A. (2005).
\newblock Designing occupancy studies: general advice and allocating survey
  effort.
\newblock {\em Journal of applied Ecology}, 42(6):1105--1114.

\bibitem[Magurran, 2013]{magurran2013measuring}
Magurran, A.~E. (2013).
\newblock {\em Measuring biological diversity}.
\newblock John Wiley \& Sons.

\bibitem[Mata et~al., 2017]{mata2017conserving}
Mata, L., Threlfall, C.~G., Williams, N.~S., Hahs, A.~K., Malipatil, M., Stork,
  N.~E., and Livesley, S.~J. (2017).
\newblock Conserving herbivorous and predatory insects in urban green spaces.
\newblock {\em Scientific reports}, 7(1):1--12.

\bibitem[Miller et~al., 2019]{miller2019recent}
Miller, D.~A., Pacifici, K., Sanderlin, J.~S., and Reich, B.~J. (2019).
\newblock The recent past and promising future for data integration methods to
  estimate species’ distributions.
\newblock {\em Methods in Ecology and Evolution}, 10(1):22--37.

\bibitem[{National Ecological Observatory Network (NEON)}, 2021]{neonData}
{National Ecological Observatory Network (NEON)} (2021).
\newblock Breeding landbird point counts (dp1.10003.001).

\bibitem[Pardieck et~al., 2020]{pardieck2020north}
Pardieck, K., Ziolkowski~Jr, D., Lutmerding, M., Aponte, V., and Hudson, M.-A.
  (2020).
\newblock North american breeding bird survey dataset 1966--2019.
\newblock {\em U.S. Geological Survey data release,
  https://doi.org/10.5066/P9J6QUF6}.

\bibitem[P{\'e}ron and Koons, 2012]{peron2012integrated}
P{\'e}ron, G. and Koons, D.~N. (2012).
\newblock Integrated modeling of communities: parasitism, competition, and
  demographic synchrony in sympatric ducks.
\newblock {\em Ecology}, 93(11):2456--2464.

\bibitem[Plummer et~al., 2006]{coda}
Plummer, M., Best, N., Cowles, K., and Vines, K. (2006).
\newblock Coda: Convergence diagnosis and output analysis for mcmc.
\newblock {\em R News}, 6(1):7--11.

\bibitem[{R Core Team}, 2020]{rSoftware}
{R Core Team} (2020).
\newblock {\em R: A Language and Environment for Statistical Computing}.
\newblock R Foundation for Statistical Computing, Vienna, Austria.

\bibitem[Rodhouse et~al., 2019]{rodhouse2019evidence}
Rodhouse, T.~J., Rodriguez, R.~M., Banner, K.~M., Ormsbee, P.~C., Barnett, J.,
  and Irvine, K.~M. (2019).
\newblock Evidence of region-wide bat population decline from long-term
  monitoring and bayesian occupancy models with empirically informed priors.
\newblock {\em Ecology and evolution}, 9(19):11078--11088.

\bibitem[Rosenberg et~al., 2019]{rosenberg2019decline}
Rosenberg, K.~V., Dokter, A.~M., Blancher, P.~J., Sauer, J.~R., Smith, A.~C.,
  Smith, P.~A., Stanton, J.~C., Panjabi, A., Helft, L., Parr, M., et~al.
  (2019).
\newblock Decline of the north american avifauna.
\newblock {\em Science}, 366(6461):120--124.

\bibitem[Royle and Dorazio, 2008]{royle2008hierarchical}
Royle, J.~A. and Dorazio, R.~M. (2008).
\newblock {\em Hierarchical modeling and inference in ecology: the analysis of
  data from populations, metapopulations and communities}.
\newblock Elsevier.

\bibitem[Schaub and Abadi, 2011]{schaub2011integrated}
Schaub, M. and Abadi, F. (2011).
\newblock Integrated population models: a novel analysis framework for deeper
  insights into population dynamics.
\newblock {\em Journal of Ornithology}, 152(1):227--237.

\bibitem[Simmonds et~al., 2020]{simmonds2020more}
Simmonds, E.~G., Jarvis, S.~G., Henrys, P.~A., Isaac, N.~J., and O'Hara, R.~B.
  (2020).
\newblock Is more data always better? a simulation study of benefits and
  limitations of integrated distribution models.
\newblock {\em Ecography}, 43(10):1413--1422.

\bibitem[Sor et~al., 2017]{sor2017effects}
Sor, R., Park, Y.-S., Boets, P., Goethals, P.~L., and Lek, S. (2017).
\newblock Effects of species prevalence on the performance of predictive
  models.
\newblock {\em Ecological Modelling}, 354:11--19.

\bibitem[Sutherland et~al., 2016]{Sutherland2016}
Sutherland, C., Brambilla, M., Pedrini, P., and Tenan, S. (2016).
\newblock {A multiregion community model for inference about geographic
  variation in species richness}.
\newblock {\em Methods in Ecology and Evolution}, 7(7):783--791.

\bibitem[Vehtari et~al., 2017]{vehtari2017practical}
Vehtari, A., Gelman, A., and Gabry, J. (2017).
\newblock Practical bayesian model evaluation using leave-one-out
  cross-validation and waic.
\newblock {\em Statistics and computing}, 27(5):1413--1432.

\bibitem[Wagner et~al., 2021]{wagner2021insect}
Wagner, D.~L., Grames, E.~M., Forister, M.~L., Berenbaum, M.~R., and Stopak, D.
  (2021).
\newblock Insect decline in the anthropocene: Death by a thousand cuts.
\newblock {\em Proceedings of the National Academy of Sciences}, 118(2).

\bibitem[Zipkin et~al., 2009]{zipkin2009impacts}
Zipkin, E.~F., DeWan, A., and Andrew~Royle, J. (2009).
\newblock Impacts of forest fragmentation on species richness: a hierarchical
  approach to community modelling.
\newblock {\em Journal of Applied Ecology}, 46(4):815--822.

\bibitem[Zipkin et~al., 2012]{zipkin2012evaluating}
Zipkin, E.~F., Grant, E. H.~C., and Fagan, W.~F. (2012).
\newblock Evaluating the predictive abilities of community occupancy models
  using auc while accounting for imperfect detection.
\newblock {\em Ecological Applications}, 22(7):1962--1972.

\bibitem[Zipkin et~al., 2010]{zipkin2010multi}
Zipkin, E.~F., Royle, J.~A., Dawson, D.~K., and Bates, S. (2010).
\newblock Multi-species occurrence models to evaluate the effects of
  conservation and management actions.
\newblock {\em Biological Conservation}, 143(2):479--484.

\bibitem[Zipkin and Saunders, 2018]{zipkin2018synthesizing}
Zipkin, E.~F. and Saunders, S.~P. (2018).
\newblock Synthesizing multiple data types for biological conservation using
  integrated population models.
\newblock {\em Biological Conservation}, 217:240--250.

\bibitem[Zipkin et~al., 2021]{zipkin2021addressing}
Zipkin, E.~F., Zylstra, E.~R., Wright, A.~D., Saunders, S.~P., Finley, A.~O.,
  Dietze, M.~C., Itter, M.~S., and Tingley, M.~W. (2021).
\newblock Addressing data integration challenges to link ecological processes
  across scales.
\newblock {\em Frontiers in Ecology and the Environment}, 19(1):30--38.

\end{thebibliography}

\end{document}